\documentclass[submission,copyright,creativecommons]{eptcs}
\providecommand{\event}{FMAS 2023} 

\usepackage{iftex}
\usepackage{graphicx}

\usepackage[colour]{lstEventB}
\lstdefinestyle{base}{
  numbers=none
}

\ifpdf
  \usepackage{underscore}         
  \usepackage[T1]{fontenc}        
\else
  \usepackage{breakurl}           
\fi

\title{Trust Modelling and Verification Using Event-B}

\author{Asieh Salehi Fathabadi
\institute{University of Southampton, United Kingdom}
\email{a-salehi-fathabadi@soton.ac.uk}
\and
Vahid Yazdanpanah
\institute{University of Southampton, United Kingdom}
\email{v.yazdanpanah@soton.ac.uk}
}
\def\titlerunning{Trust Verification using Event-B}
\def\authorrunning{A. Salehi and V. Yazdanpanah}
\usepackage{color}

\usepackage{amsthm}

\theoremstyle{definition}
\newtheorem{definition}{Definition}

\theoremstyle{remark}

\begin{document}
\maketitle

\begin{abstract}



Trust is a crucial component in collaborative multiagent systems (MAS) involving humans and autonomous AI agents. Rather than assuming trust based on past system behaviours, it is important to formally verify trust by modelling the current state and capabilities of agents. We argue for verifying actual trust relations based on agents' abilities to deliver intended outcomes in specific contexts. To enable reasoning about different notions of trust, we propose using the refinement-based formal method Event-B. Refinement allows progressively introducing new aspects of trust - from abstract to concrete models incorporating knowledge and runtime states. We demonstrate modelling three trust concepts and verifying associated trust  properties in MAS. The formal, correctness-by-construction approach allows to deduce guarantees about trustworthy autonomy in human-AI partnerships. Overall, our contribution facilitates rigorous verification of trust in multiagent systems.
\end{abstract}

\section{Introduction}

Trust is a crucial contextual concept in multiagent systems (MAS), representing the cognitive state of a trustor towards a trustee~\cite{castelfranchi2020trust}. While there are various accounts of trust, in this work, we focus on trust with respect to accomplishing tasks. While trust modelling in MAS has historically relied on reasoning about past behaviours\cite{ramchurn2004trust}, recent work emphasises integrating current context rather than fully depending on history. This involves verifying what agents can actually deliver based on their present capabilities, beyond reputations. We argue for complementing offline safety assurances with online trust verification for autonomous systems. Consider an autonomous delivery vehicle (ADV) tasked with transporting goods. Offline verification during design suffices for basic safety and whether the ADV is reliable in general (regardless of their current state and how they can perform in the context). However, assessing trust online for a particular delivery also requires checking the ADV's abilities given its current battery, payload etc. against user requirements.

Trust modelling in MAS, and what we introduce as ``actual trust'', entails representing different aspects like agents' abilities, knowledge and commitments. To model such a multidimensional notion, refinement techniques like Event-B~\cite{abrial10} allow correct-by-construction modelling~\cite{Lanoix08,Gao07}. 
Our key contribution is a refinement-based approach that supports formally verifying various  trust concepts. We demonstrate formally modelling three trust notions relating to agent abilities, knowledge and commitments. The automated consistency guarantees complement offline assurance for trustworthy autonomy and human-AI partnerships~\cite{ramchurn21,stein2023citizen}. This work is an  initial step on modelling trust using Event-B's refinement strategy that practically enables step-wise verification of actual trust between agents  in autonomous systems. 


\section{Actual Trust: Power, Knowledge, and Commitments}\label{sec:analysis}

In modelling and reasoning about trust, it is key to distinguish what an agent may rely on due to past behaviour of another agent and their \textit{typical} behaviour from what in a given situation agents are \textit{actually} able to deliver. While the former category of trusting has a retrospective view, and uses history to reason about trust~\cite{ramchurn2004trust}, the latter form of trust is to reason about what the other agent can \textit{actually} deliver and has basis in what is known in the theory of causality as \textit{actual causality}~\cite{halpern2016actual}. In this work, we focus on the latter notion, refer to it as  \textit{actual trust} and understand it as a relational notion between two agents or agent groups $i$ (as the trustor) and $j$ (as the trustee) and say in a particular multiagent system $M$, $i$ trusts $j$ with respect to task $t$ only if $i$ is able to verify that $j$ is able and committed to deliver $t$. 
To model and verify our notion of actual trust, as knowledge of another agents' ability and commitment to ensure a particular task, it is important to highlight how it relates to its key components conceptually.

\textbf{Trusting for ability to materialise eventualities:} In contrast to purely history-oriented perspectives to trust, that look at the history and trust an agent to behave similar to its past behaviour, we deem that trusting needs to be fine-tuned based on the current state of the system and   actions agents are able to execute and what agents intend to deliver. For instance, even if an ADV was successful in former deliveries, it may be suffering from a low battery now and unable to deliver tasks. So, one should fine tune trust in the agent's power to deliver based on the current situation.

\textbf{Trust as an epistemic state:} We understand  actual trust as an epistemic notion meaning that it is essentially about knowledge of the trustor on how another agent relates to a particular event. Recalling the running example, the user needs to reason about abilities of an ADV, consider its publicly announced intentions, and verify if the ADV can be trusted for a particular delivery. This form of trust allows specification of trust in different contexts and for different types requirements and knowledge levels. For instance, a given ADV $j$ may be seen as ``trusted for delivering $5kg$ of groceries'' but this trust may not extend when it comes to passenger pickup.. 

\textbf{Public commitments as a proxy to intentions:} When we are dealing with autonomous AI agents, we need to consider that being able to deliver a task fundamentally differs  from delivering the task. Imagine that an ADV $v$ with a full battery and ability to deliver some goods is located relatively close to an agent $i$ with a delivery task $t$. In this case, $i$ can't simply assume that $v$ can be trusted for delivering $t$ as it may be already committed  to deliver tasks other than $t$ or is in the middle of other plan executions. To handle this, we use notion of publicly-announced commitments as a proxy to model what agents intend to bring about.\footnote{Note that assuming full access to agents' intentions is against separation of concerns, privacy, and encapsulation as key design principles in safe and responsible AI and software development.}





\section{Refinement-Based Trust Formal Modelling and Verification}
\label{sec:model}

\begin{mdframed}
\textbf{Background knowledge:}
\EventB~\cite{abrial10} is a refinement-based formal method for system development. The mathematical language of \EventB is based on set theory and first order logic\footnote{Please refer to Event-B Language user manual~\url{https://wiki.event-b.org/index.php/Event-B_Language} for extra support to understand the presented model.}. An \EventB model consists of two parts: \textit{contexts} for static data and \textit{machines} for dynamic behaviour. An \EventB model is constructed by making progressive refinements starting from an initial abstract model which may have more general behaviours and gradually introducing more detail that constrains the behaviour towards the desired system. Each refinement step is verified to be a valid refinement of the previous step. 
\end{mdframed}

\begin{figure}[htbp]
	\centering
	\includegraphics[width=0.6\textwidth]{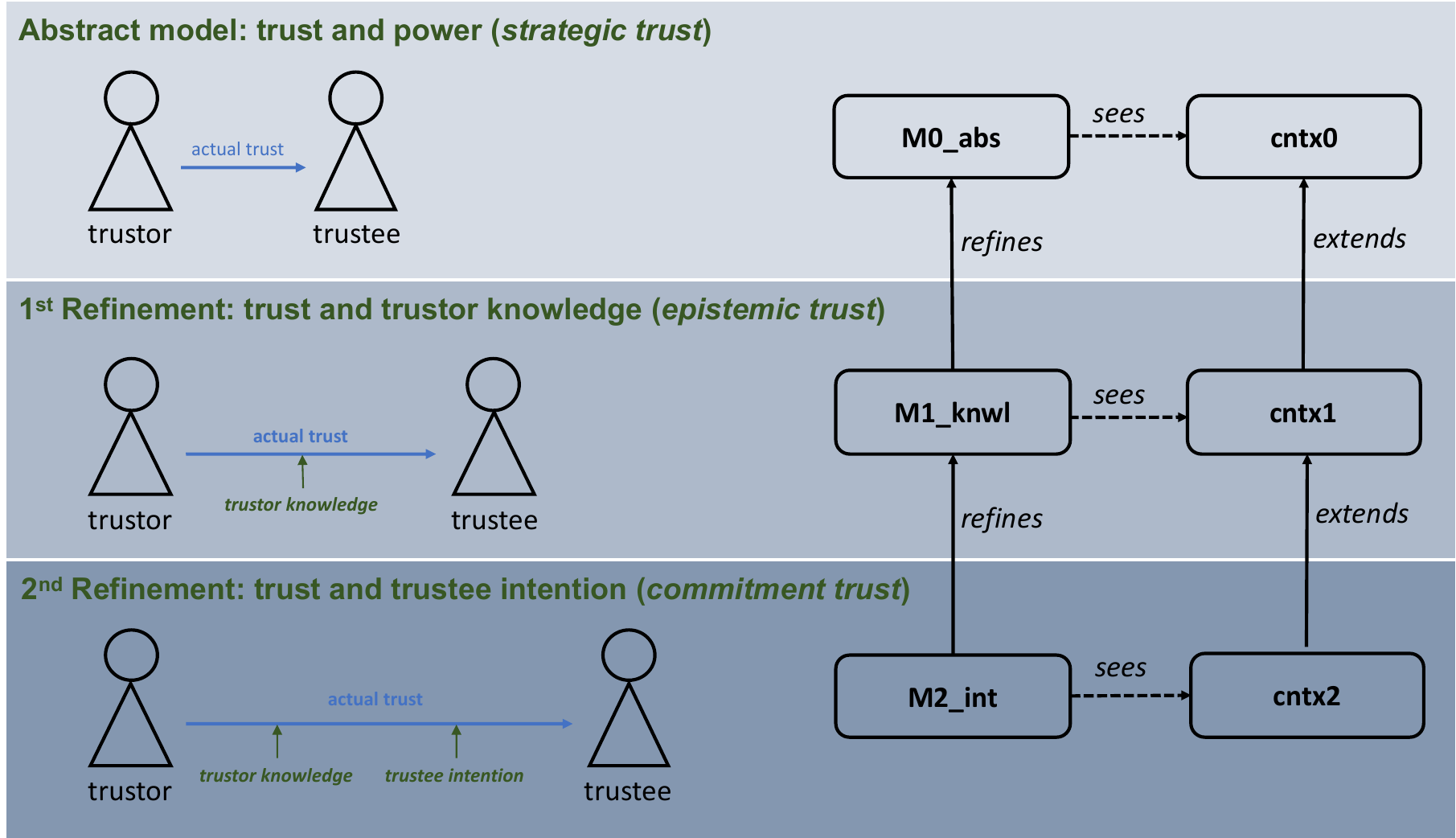}
	\caption{Trust Modelling and the Refinement Strategy}
	\label{Figure:refStra}
\end{figure}

Benefiting from refinement technique in building Event-B formal model, instead of one single-layer complex design model of system, we propose to gradually introduce different concepts of actual trust through refinement steps.
Figure~\ref{Figure:refStra} presents our vision idea of applying refinement-based development to model actual trust in autonomous systems~\footnote{Note that because of space limitation, the Event-B model of trust is not fully presented here. And for simplicity, in purpose of demonstrating the vision idea, we model trust in its simplest definition.}. 
Left side illustrates the trust relationship between trustor and trustee, while right side presents the structure of our proposed Event-B formal model, including three levels of refinements: machines ($M0$, $M1$ and $M2$) and associated contexts ($cntx0$, $cntx1$ and $cntx2$).

In line with trust key components and first-order building blocks presented in Section~\ref{sec:analysis}, starting from the top level, trust is first modelled as an abstract relationship between two agents, trustor and trustee, \textit{M0_abs}; followed by first refinement level where \textit{trustor knowledge} is introduced, \textit{M1_knwl}. Then \textit{trustee intention} is introduced in a further refinement level, \textit{M2_int}.

\vspace{-0.2cm}
\subsection{Modelling trust in Event-B}
Agents and tasks are defined as a set in the context \textit{cntx0}, which is partitioned to two sub-sets: trustors and trustees:

\begin{mdframed}
\textbf{Background knowledge:} An Event-B context contains carrier sets |s|, constants |c|, and axioms |A(c)| that constrain the carrier sets and constants.\end{mdframed}

\begin{EventBcode}
CONTEXT  cntx0
SETS   AGENTS    TASKS    CONSTANTS   trustors    trustees
AXIOMS   axm1   :   	trustors ⊆ AGENTS   // Definition: Subset ⊆
                       axm2   :   	trustees ⊆ AGENTS
                       axm3   :   	partition (AGENTS, trustors, trustees)
\end{EventBcode}

\begin{definition}
    Abstract (strategic) trust: agent $i$ weakly (abstraction) trusts $j$ regarding task $t$ if $j$ has an action or a sequence of actions available to it to ensure $e$. We are operating in a cooperative setting, hence assuming that agents share information and have perfect knowledge of themselves as well as other agents' abilities. We define trust in terms of the ability to deliver $t$. 
\end{definition}

Trust is modelled as a three-dimension relation variable between a trustor, a set of trustees and a task, in the abstract machine.
In \textit{M0_abs}, an invariant, $inv1$, is specifying the \textit{agent_task} variable as a function between a subset of trustees and a single task, indicating the task that can be delivered by a subset of trustee agents. And $inv2$ is specifying the \textit{trustor_trustee_task} variable indicating the relation between a trustor and a pair of \textit{agent_task}.

\begin{mdframed}
\textbf{Background knowledge:} An Event-B machine contains variables |v|, invariant predicates |I(v)| that constrain the variables, and events. In Event-B, a machine corresponds to a transition system where \emph{variables} represent the states and \emph{events} specify the transitions.
\end{mdframed}

\begin{EventBcode}
    @inv1: agent_task ∈ ℙ(trustees) ⇸ tasks    // specifies which agents are able to deliver which task
    // Definition: Powerset: ℙ(S) = {s | s ⊆ S}
    // Definition: A function (agent_task) is a relation with the restriction that each element of the domain (ℙ(trustees)) is related to a unique element in the range (tasks); a many to one mapping
    // Definition: Set membership ∈ 
    @inv2: 	trustor_trustee_task ∈ trustors ⇸ agent_task	// specifies set of triples i ↦ (j ↦ t), when agent i can trust a set of agents j to deliver task t 
\end{EventBcode}

The event $trust$ is adding a new triple to the \textit{trustor_trustee_task} variable, $act1$. While $grd1-3$ is checking the type of the event parameters, $grd4$ is indicating the above definition, ensuring $j$ is able to deliver task $t$; guards $grd5-8$ are described later.

\begin{mdframed}
\textbf{Background knowledge:} An event in a machine, comprises a guard denoting its enabling-condition and an action describing how the variables are modified when the event is executed.  In general, an event |e| has the following form, where |t| are the event parameters, |G(t, v)| is the guard of the event, and |v := E(t, v)| is the action of the event: |e == any t where G(t,v) then v := E(t,v) end|
\end{mdframed}

\begin{EventBcode}
    	event trust   any  i  j  t
    	where   @grd1   :   	i ∈ trustors
                      @grd2   :   	j ∈ ℙ(trustees)
                      @grd3   :   	t ∈ tasks
                      @grd4   :     t ∈ agent_task[{j}]   // j is able to deliver task t
                      // Definition: relational image: r[S] = {y | ∃x·x ∈ S ∧ x |-> y ∈ r} where S is a set
                      @grd5   :   	i ∉ j   // to preserve inv3
                      // Definition: Set non-membership ∉ 
                      @grd6   :   	j ≠ ∅   // abstract guard to preserve inv4
            @grd7   :   	j ⊆ knowledge[{i}]   // refining guard to preserve inv4
            @grd8   :   	commitments[i ↦ (j ↦ t)] = {TRUE}   // refining guard to preserve inv4
    	then     @act1: trustor_trustee_task ≔ trustor_trustee_task ∪ {i ↦ (j ↦ t)}
     // Definition: Union ∪ 
\end{EventBcode}

\vspace{-0.5cm}
\paragraph{Running example:} For instance, for an agent $i$ and an ADV $j$ and task of “delivering 5kg of groceries”, $i$ can trust $j$ only if “delivering 5kg of groceries” is within the allocated tasks to $j$: $grd3$.
Then, $act1$ will add a new triple of $(i, j, t)$ to the variable set \textit{trustor_trustee_task}. 

\vspace{-0.2cm}
\subsection{Modelling verifiable trust properties}
To propose the idea of formal verification of properties of trust in autonomous systems, here we present two invariants, specifying two fundamental trust properties.
\textit{inv3} is specifying that an agent $i$ would not trust itself to deliver a specific task $t$.
And \textit{inv4} is specifying avoiding trust deadlock, that for each trustor $i$ and task $t$, there is always a non-empty subset of trustees $j$ that can deliver $t$.

\begin{EventBcode}
    @inv3: ∀ i , j· i ∈ trustors ∧ j ∈ ℙ(trustees) ∧ i ∈ dom(trustor_trustee_task) ⇒ i ∉ j	
    // Definition: Conjunction ∧ , Universal quantification ∀ , Implication ⇒
    // Definition: Domain: dom(r) ∀r·r ∈ S ↔ T ⇒ dom(r) = {x·(∃y·x 7→ y ∈ r)} where S and T are sets
    @inv4: ∀ i , t · i ∈ trustors ∧ t ∈ tasks ⇒ (∃ j · j ∈ ℙ(trustees) ∧  j ≠ ∅)
    // Definition: Existential quantification ∃
\end{EventBcode}

\vspace{-0.2cm}
\subsection{Verifying trust properties}

\begin{mdframed}
\textbf{Background knowledge:} 
\EventB is supported by the Rodin tool set~\cite{abrialButler10}, an extensible open source toolkit which includes facilities for modelling, verifying the consistency of models using theorem proving and model checking techniques, and validating models with simulation-based approaches.
\end{mdframed}

One of the generated proof obligations (PO) for an Event-B model, is ''invariant preservation'': 

|e / v / INV| (where $e$ is the event name, and $v$ is the invariant name)

|INV| PO ensures that the property specified in the invariant |INV| is preserved by event |e|.
To preserve the trust properties defined in $inv3$ and $inv4$, the event $trust$ is guarded by $grd5$ and $grd6$, see above. 
Two POs |trust/inv3/INV| and |trust/inv4/INV| are generated and automatically discharged by Rodin tool.

\vspace{-0.2cm}
\subsection{Refining trust} Next, we introduce the refined notion of epistemic trust in which agents' knowledge is integrated. 
\begin{definition}
    Refined (epistemic) trust: for a stronger notion of trust we require a variable of $knowledge$ specifying the knowledge relationship between two agents $i, j$, indicating whether $i$ is fully aware of $j$'s abilities.
\end{definition}

Refining model \textit{M1_knwl} introduces the $knowledge$ variable to model the knowledge of trustors about trustees:

\begin{EventBcode}
    @inv1: knowledge ∈ trustors ↔ trustees 
    // Definition: A relation (knowledge) is a set of ordered pairs; a many to many mapping.
\end{EventBcode}

\vspace{-0.5cm}
\paragraph{Running example:} For instance, a given ADV $j$ may be seen as ``trusted for delivering $5kg$ of groceries''  by an agent $i$ who is fully aware of $j$'s abilities but not by agent $v$ who is not 
aware of $j$ and  that $j$ has the capacity to ensure $t$. 

\begin{definition}
    Refined (commitment) trust: for a stronger notion of trust we require a variable of $commitments$ specifying a function that takes a trust triple $(i, j, t)$ and determines whether agent $j$ is committed to deliver task $t$ for agent $i$. We refine the trust model, not only in terms of the ability, but also the commitment to deliver $t$. 
\end{definition}

And refining model \textit{M2_int} introduces the $commitment$ variable to model the intention of trustees to deliver the associate task (for simplicity in this paper, we model commitment as a Boolean indicating whether an agent(s) as trustee intends to deliver the associated task or not):

\begin{EventBcode}
    @inv1: commitments ∈ trustor_trustee_task → BOOL
\end{EventBcode}

$inv4$ is refined to include the knowledge property in \textit{M1_knwl} and commitment specification in \textit{M2_int}:

\begin{EventBcode}
    @inv4: ∀ i , t · i ∈ trustors ∧ t ∈ tasks ⇒ 
                  (∃ j · j ∈ ℙ(trustees) ∧ j ≠ ∅ ∧ (j ↦ t) ∈ agent_task ∧ 
                  j ⊆ knowledge[{i}] ∧ commitments[i ↦ (j ↦ t)] = {TRUE} ∧ 
                  i ↦ (j ↦ t) ∈ trustor_trustee_task)
\end{EventBcode}

And refining event $trust$ includes extra guards $grd7$ and $grd8$ to preserve $inv4$, see above. Not providing these guards results in failed generated |INV| POs. 

\vspace{-0.5cm}
\paragraph{Running example:} For instance, for an agent $i$ and an ADV $j$ and task of “delivering 5kg of groceries”, $i$ can trust $j$ only if “delivering 5kg of groceries” is within the allocated tasks to $j$: $grd3$ (verified in the abstract machine), and $i$ is fully aware of j’s abilities: $grd7$ (verified in the machine \textit{M1_knwl}) and $j$ is committed to deliver 5kg of groceries to $i$: $grd8$ (verified in the machine \textit{M2_int}).

\vspace{-0.2cm}
\paragraph{Model checking trust properties:} 
The presented Event-B model can be model checked by instantiating the context elements, for example for the ADV system. Also the scenario checker integrated in Rodin can demonstrate difference scenarios of the desire system. Due to the concise  nature of this work and space limitation, we are unable to include the model checking experience.

\vspace{-0.2cm}


\section{Concluding Remarks and Future Directions}
\label{sec:conclusion}

The step-wise refinement approach presented in this paper, demonstrates three notions of actual trust, and two verifiable trust properties. The model can simply refined to include more notions and properties. This paper elaborates on how the autonomous system research can benefit from refinement-based formal methods in terms of modelling trust. The abstraction technique aids the modelling and verification process in step-wise manner, where instead of a single complex model, the formal model is gradually built through refinement levels, hence easier to be understood and proved. Also the Event-B formal method provides the verification techniques (theorem proving and ProB model checking~\cite{Leuschel08}) in each refinement level, to ensure the trustworthiness of autonomous systems.

\textbf{Contributions to  Autonomous Systems (AS):}
In AS, replacing human decision-making with machine decision-making results in challenges associated with stakeholders’ trust. Trustworthiness of an AS is key to its wide-spread adoption by society.
To develop a trusted AS, it is important to understand how different stakeholders perceive an AS as trusted, and how the context of application affects their perceptions.
The translation of trust issues into formalised solutions is challenging due to trust dynamics. In this work, we try to advance in this direction by utilising the ability of Event-B as a refinement-based formal method to manage the lack of information when modelling trust in multi-agent systems. High-level model aids to abstract away the uncertain/unknown trust specifications. 
We introduced the notion of actual trust versus statistical trust, toward trusting to an AS due to its safety checks, like, inherent uncertainties in the environment, diversity in the requirements and needs of different users and contexts of application. We formalised the notion of actual trust using Event-B formal modelling followed by verifying the safety properties of it. The actual trust notions is modelled and verified in three levels: strategic trust, epistemic trust and commitment trust.


  


 



\textbf{Future directions:}
This formal modelling and verification approach for trust in autonomous systems can be extended in several directions. One avenue is via integrating Event-B models with Alternating-Time Temporal Logic~\cite{zhu2023fairness} to allow more expressive temporal specifications and model checking of trust properties, e.g., in the context of connected mobility systems. Further research can also investigate gradation of trust (as a quantifiable notion) and formally relating trust and neighbouring notions in multiagent settings such as responsibility~\cite{yazdanpanah2016quantified}. Quantifying trust based on strategy lengths and information-theoretic notions may also complement the  approach pursued here. Overall, rigorous formal methods can provide significant assurances about trustworthy autonomy and human-AI partnerships, especially for safety-critical applications.

\textit{Acknowledgements:} This work was supported by the UK Engineering and Physical Sciences Research Council (EPSRC) through a Turing AI Fellowship (EP/V022067/1) on Citizen-Centric AI Systems (\url{https://ccais.ac.uk/}) and
the UKRI Trustworthy Autonomous Systems Hub (EP/V00784X/1).  

\nocite{*}
\bibliographystyle{eptcs}
\bibliography{ref}

\begin{thebibliography}{10}
\providecommand{\bibitemdeclare}[2]{}
\providecommand{\surnamestart}{}
\providecommand{\surnameend}{}
\providecommand{\urlprefix}{Available at }
\providecommand{\url}[1]{\texttt{#1}}
\providecommand{\href}[2]{\texttt{#2}}
\providecommand{\urlalt}[2]{\href{#1}{#2}}
\providecommand{\doi}[1]{doi:\urlalt{https://doi.org/#1}{#1}}
\providecommand{\eprint}[1]{arXiv:\urlalt{https://arxiv.org/abs/#1}{#1}}
\providecommand{\bibinfo}[2]{#2}

\bibitemdeclare{book}{abrial10}
\bibitem{abrial10}
\bibinfo{author}{J-R. \surnamestart Abrial\surnameend} (\bibinfo{year}{2010}):
  \emph{\bibinfo{title}{Modeling in {Event-B}: System and Software
  Engineering}}.
\newblock \bibinfo{publisher}{Cambridge University Press},
  \doi{10.1017/S0956796812000081}.

\bibitemdeclare{article}{abrialButler10}
\bibitem{abrialButler10}
\bibinfo{author}{J-R \surnamestart Abrial\surnameend},
  \bibinfo{author}{M.~\surnamestart Butler\surnameend},
  \bibinfo{author}{S.~\surnamestart Hallerstede\surnameend},
  \bibinfo{author}{T.S. \surnamestart Hoang\surnameend},
  \bibinfo{author}{F.~\surnamestart Mehta\surnameend} \&
  \bibinfo{author}{L.~\surnamestart Voisin\surnameend} (\bibinfo{year}{2010}):
  \emph{\bibinfo{title}{{Rodin}: An Open Toolset for Modelling and Reasoning in
  {Event-B}}}.
\newblock {\slshape \bibinfo{journal}{Software Tools for Technology Transfer}}
  \bibinfo{volume}{12}(\bibinfo{number}{6}), pp. \bibinfo{pages}{447--466},
  \doi{10.1007/s10009-010-0145-y}.

\bibitemdeclare{article}{castelfranchi2020trust}
\bibitem{castelfranchi2020trust}
\bibinfo{author}{Cristiano \surnamestart Castelfranchi\surnameend} \&
  \bibinfo{author}{Rino \surnamestart Falcone\surnameend}
  (\bibinfo{year}{2020}): \emph{\bibinfo{title}{Trust: Perspectives in
  cognitive science}}.
\newblock {\slshape \bibinfo{journal}{The Routledge Handbook of Trust and
  Philosophy}}, pp. \bibinfo{pages}{214--228}, \doi{10.4324/9781315542294-17}.

\bibitemdeclare{article}{dastani2023responsibility}
\bibitem{dastani2023responsibility}
\bibinfo{author}{Mehdi \surnamestart Dastani\surnameend} \&
  \bibinfo{author}{Vahid \surnamestart Yazdanpanah\surnameend}
  (\bibinfo{year}{2023}): \emph{\bibinfo{title}{Responsibility of AI systems}}.
\newblock {\slshape \bibinfo{journal}{Ai \& Society}}
  \bibinfo{volume}{38}(\bibinfo{number}{2}), pp. \bibinfo{pages}{843--852},
  \doi{10.1007/s00146-022-01481-4}.

\bibitemdeclare{article}{Gao07}
\bibitem{Gao07}
\bibinfo{author}{Hang-Jiang \surnamestart Gao\surnameend},
  \bibinfo{author}{Zheng \surnamestart Qin\surnameend}, \bibinfo{author}{Lei
  \surnamestart Lu\surnameend}, \bibinfo{author}{Li-Ping \surnamestart
  Shao\surnameend} \& \bibinfo{author}{Xing-Chen \surnamestart Heng\surnameend}
  (\bibinfo{year}{2007}): \emph{\bibinfo{title}{Formal specification and proof
  of multi-agent applications using event b}}.
\newblock {\slshape \bibinfo{journal}{Information Technology Journal}}
  \bibinfo{volume}{6}(\bibinfo{number}{7}), pp. \bibinfo{pages}{1181--1189},
  \doi{10.3923/itj.2007.1181.1189}.

\bibitemdeclare{book}{halpern2016actual}
\bibitem{halpern2016actual}
\bibinfo{author}{Joseph~Y \surnamestart Halpern\surnameend}
  (\bibinfo{year}{2016}): \emph{\bibinfo{title}{Actual causality}}.
\newblock \bibinfo{publisher}{MiT Press},
  \doi{10.7551/mitpress/10809.001.0001}.

\bibitemdeclare{inproceedings}{Lanoix08}
\bibitem{Lanoix08}
\bibinfo{author}{Arnaud \surnamestart Lanoix\surnameend}
  (\bibinfo{year}{2008}): \emph{\bibinfo{title}{Event-B Specification of a
  Situated Multi-Agent System: Study of a Platoon of Vehicles}}.
\newblock In: {\slshape \bibinfo{booktitle}{Second {IEEE/IFIP} International
  Symposium on Theoretical Aspects of Software Engineering, {TASE} 2008, June
  17-19, 2008, Nanjing, China}}, \bibinfo{publisher}{{IEEE} Computer Society},
  pp. \bibinfo{pages}{297--304}, \doi{10.1109/TASE.2008.39}.

\bibitemdeclare{article}{Leuschel08}
\bibitem{Leuschel08}
\bibinfo{author}{Michael \surnamestart Leuschel\surnameend} \&
  \bibinfo{author}{Michael \surnamestart Butler\surnameend}
  (\bibinfo{year}{2008}): \emph{\bibinfo{title}{{ProB}: {An} Automated Analysis
  Toolset for the {B} Method}}.
\newblock {\slshape \bibinfo{journal}{Software Tools for Technology Transfer
  (STTT)}} \bibinfo{volume}{10}(\bibinfo{number}{2}), pp.
  \bibinfo{pages}{185--203}, \doi{10.1007/s10009-007-0063-9}.

\bibitemdeclare{article}{ramchurn2004trust}
\bibitem{ramchurn2004trust}
\bibinfo{author}{Sarvapali~D \surnamestart Ramchurn\surnameend},
  \bibinfo{author}{Dong \surnamestart Huynh\surnameend} \&
  \bibinfo{author}{Nicholas~R \surnamestart Jennings\surnameend}
  (\bibinfo{year}{2004}): \emph{\bibinfo{title}{Trust in multi-agent systems}}.
\newblock {\slshape \bibinfo{journal}{The knowledge engineering review}}
  \bibinfo{volume}{19}(\bibinfo{number}{1}), pp. \bibinfo{pages}{1--25},
  \doi{10.1017/S0269888904000116}.

\bibitemdeclare{article}{ramchurn21}
\bibitem{ramchurn21}
\bibinfo{author}{Sarvapali~D \surnamestart Ramchurn\surnameend},
  \bibinfo{author}{Sebastian \surnamestart Stein\surnameend} \&
  \bibinfo{author}{Nicholas~R \surnamestart Jennings\surnameend}
  (\bibinfo{year}{2021}): \emph{\bibinfo{title}{Trustworthy human-AI
  partnerships}}.
\newblock {\slshape \bibinfo{journal}{Iscience}}
  \bibinfo{volume}{24}(\bibinfo{number}{8}), p. \bibinfo{pages}{102891},
  \doi{10.1016/j.isci.2021.102891}.

\bibitemdeclare{inproceedings}{stein2023citizen}
\bibitem{stein2023citizen}
\bibinfo{author}{Sebastian \surnamestart Stein\surnameend} \&
  \bibinfo{author}{Vahid \surnamestart Yazdanpanah\surnameend}
  (\bibinfo{year}{2023}): \emph{\bibinfo{title}{Citizen-Centric Multiagent
  Systems}}.
\newblock In: {\slshape \bibinfo{booktitle}{Proceedings of the 2023
  International Conference on Autonomous Agents and Multiagent Systems}}, pp.
  \bibinfo{pages}{1802--1807}.
\newblock \urlprefix\url{https://dl.acm.org/doi/10.5555/3545946.3598843}.

\bibitemdeclare{inproceedings}{yazdanpanah2016quantified}
\bibitem{yazdanpanah2016quantified}
\bibinfo{author}{Vahid \surnamestart Yazdanpanah\surnameend} \&
  \bibinfo{author}{Mehdi \surnamestart Dastani\surnameend}
  (\bibinfo{year}{2016}): \emph{\bibinfo{title}{Quantified degrees of group
  responsibility}}.
\newblock In: {\slshape \bibinfo{booktitle}{Coordination, Organizations,
  Institutions, and Norms in Agent Systems}}, \bibinfo{organization}{Springer},
  pp. \bibinfo{pages}{418--436}, \doi{10.1007/978-3-319-42691-4\_23}.

\bibitemdeclare{article}{zhu2023fairness}
\bibitem{zhu2023fairness}
\bibinfo{author}{Chenyang \surnamestart Zhu\surnameend},
  \bibinfo{author}{Michael \surnamestart Butler\surnameend},
  \bibinfo{author}{Corina \surnamestart Cirstea\surnameend} \&
  \bibinfo{author}{Thai~Son \surnamestart Hoang\surnameend}
  (\bibinfo{year}{2023}): \emph{\bibinfo{title}{A fairness-based refinement
  strategy to transform liveness properties in Event-B models}}.
\newblock {\slshape \bibinfo{journal}{Science of Computer Programming}}
  \bibinfo{volume}{225}, p. \bibinfo{pages}{102907},
  \doi{10.1016/j.scico.2022.102907}.

\end{thebibliography}
\end{document}